\documentclass{article}
\usepackage{spconf,amsmath,graphicx}

\usepackage{spconf,amsmath,graphicx}
\usepackage{subcaption}
\usepackage{booktabs}
\usepackage{multirow}
\usepackage{adjustbox}
\usepackage{graphicx}
\usepackage{lipsum}
\usepackage{verbatim}
\usepackage{url}
\usepackage{amssymb}

\title{Transduce and Speak: Neural Transducer for Text-to-Speech with Semantic Token Prediction}
\name{Minchan Kim, Myeonghun Jeong, Byoung Jin Choi, Dongjune Lee, Nam Soo Kim}
\address{Department of Electrical and Computer Engineering and INMC, \\Seoul National University, Seoul, South Korea}
\copyrightnotice{979-8-3503-0689-7/23/\$31.00~\copyright2023 IEEE}
\begin{document}
%
\maketitle
\begin{abstract}
We introduce a text-to-speech~(TTS) framework based on a neural transducer. We use discretized semantic tokens acquired from wav2vec2.0 embeddings, which makes it easy to adopt a neural transducer for the TTS framework enjoying its monotonic alignment constraints. The proposed model first generates aligned semantic tokens using the neural transducer, then synthesizes a speech sample from the semantic tokens using a non-autoregressive~(NAR) speech generator. This decoupled framework alleviates the training complexity of TTS and allows each stage to focus on 1) linguistic and alignment modeling and 2) fine-grained acoustic modeling, respectively. Experimental results on the zero-shot adaptive TTS show that the proposed model exceeds the baselines in speech quality and speaker similarity via objective and subjective measures. We also investigate the inference speed and prosody controllability of our proposed model, showing the potential of the neural transducer for TTS frameworks.
\end{abstract}
\begin{keywords}
speech synthesis, neural transducer, zero-shot adaptive text-to-speech, speech representation
\end{keywords}
\section{Introduction}
\label{sec:intro}
Neural Text-to-Speech~(TTS) systems face two main problems: alignment modeling and one-to-many sequence generation. Because of the strong inductive bias that text and speech are monotonically aligned, learning and representing the alignment efficiently are important parts of designing a TTS system. For example, autoregressive~(AR) TTS models~\cite{shen2018natural, li2019neural, kharitonov2023speak} find alignments by themselves using attention mechanisms. On the contrary, non-autoregressive~(NAR) TTS family~\cite{ren2020fastspeech, kim2021conditional, popov2021grad} uses external alignment search algorithms~\cite{mcauliffe2017montreal, kim2020glow} and phoneme-wise duration predictors for length regulation. As a sentence can be spoken in various ways, representing and controlling the diversity of speech are also crucial issues for TTS. Numerous factors, including speaker, prosody, and recording conditions, can induce this one-to-many problem. Recent TTS models use deep generative models or conditioning information such as reference speech to cover this problem. Although these two issues can be treated independently, we assume that they are highly interrelated each other. For example, when the diversity of the data gets larger, especially in the case of zero-shot adaptive TTS, it becomes harder to learn the exact alignment, and the misalignment can make training unstable, which results in degraded sample quality and insufficient controllability. In this perspective, to alleviate this complexity, we decompose the TTS pipeline into alignment modeling and fine-grained acoustic modeling, and then solve each part with carefully designed methods.

Meanwhile, one of the sophisticated sequence generation models with monotonic alignment is the neural transducer~(e.g. RNN-T)~\cite{graves2012sequence}. As a sequence generator, the neural transducer finds the alignment using a special $\langle blank\rangle$ token and learns the marginal conditional likelihood of sequences of all the possible monotonic alignments. Although these properties seem suitable for the TTS objective as for automatic speech recognition~(ASR), it is difficult to directly adopt a neural transducer for TTS because TTS has continuous output space, while the neural transducer is suitable for discrete sequences. Among previous works~\cite{yasuda2019initial, chen2021speech} which leveraged the neural transducer, SSNT-TTS~\cite{yasuda2019initial} calculates the joint probability of mel-spectrogram and alignment, while Speech-T~\cite{chen2021speech} proposed a lazy forward algorithm which decouples aligning and mel-spectrogram prediction. However, these regression-based approaches still have limitations as they cannot represent one-to-many distributions of speech. Therefore, these previous works only explored single-speaker TTS with limited data complexity.

In this paper, we propose a TTS model which consists of two stages: text-to-token~(T2T) and token-to-speech~(T2S) stage. The token, named semantic token, indicates the index of $k$-means clustering on wav2vec2.0 embeddings~\cite{baevski2020wav2vec}. As the semantic token has disentangled semantic information, we can separately focus on 1) semantic and alignment modeling in the T2T stage and 2) acoustic diversity modeling in the T2S stage. Especially for the T2T stage, we employ a neural transducer. As semantic tokens are in a discrete domain, we can easily take advantage of the neural transducer, including its monotonic alignment constraints. This allows us more efficient and robust alignment modeling. For the T2S stage, we use a VITS-based architecture~\cite{kim2021conditional} for high-fidelity and fast speech generation. To control the paralinguistic attributes such as prosody and speaker, we condition a reference speech on both stages. The references speech respectively controls alignment-related prosody~(e.g. speech rate) in the T2T stage and fine-grained acoustic details~(e.g. speaker, recording condition) in the T2S stage. Our experiment results on zero-shot adaptive TTS demonstrate that the proposed model outperforms the baselines regarding intelligibility, naturalness, and zero-shot speaker adaptation. In addition, we found that the proposed model generates speech much faster than in real-time and is able to control prosody using the reference speech of the neural transducer. We provide the samples on our demo page: https://gannnn123.github.io/transduce-and-speak

\section{Proposed Method}
\label{sec:pagestyle}
This section introduces the novel two-stage TTS framework with text-to-token~(T2T) and token-to-speech~(T2S) stages. In the T2T stage, text inputs are converted into a sequence of semantic tokens, and semantic tokens are converted back into a speech sample in the T2S stage. Semantic tokens are obtained from representations of the wav2vec2.0 model. Similar to \cite{kim2022transfer, kharitonov2023speak}, we conduct $k$-means clustering on wav2vec2.0 embeddings and use the cluster index sequence as the semantic tokens. The details of each stage are described in sections 3.1 and 3.2.

\subsection{Text-to-Token~(T2T)}
We employ a neural transducer for the T2T stage. The neural transducer learns the alignments and encodes high-level semantic information from text sequences. Exploiting the semantic tokens as a target helps to overcome the limitations of adopting a neural transducer for TTS. Quantization by $k$-means clustering makes the output space discrete, which enables the adoption of a neural transducer without using complex schemes. In addition, using semantic tokens reduces the data complexity induced by fine-grained speech conditions, which makes it easier for the neural transducer to learn the alignment and semantic information.

\begin{figure}[]
 \centering
 \includegraphics[width=0.75\columnwidth]{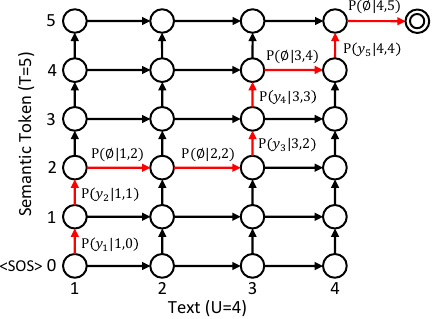}
 \caption{The output probability lattice for the neural transducer. The red path shows an example of monotonic alignment between text and semantic tokens.}
 \vspace{-0.2cm}
 \label{fig:t-SNE}
 \end{figure}
 
\subsubsection{Neural transducer for semantic token prediction}
We follow the formulation of a neural transducer for our T2T model. Let's denote the text sequence $X=\{x_u\}^{U}_{u=1}$ and the target semantic token sequence $Y=\{y_t\}^{T}_{t=0}$,  where $U$ and $T$ are lengths of text and semantic token sequences respectively and $y_0$ corresponds to the $\langle SOS\rangle$ token. The neural transducer learns conditional likelihood~$P(Y|X)$ using the lattice~$\{\alpha(u,t)|{1\le u\le {U}}, {0\le t\le {T}}\}$. As depicted in Fig.~1, each node $\alpha(u,t)$ represents the probability of emitting first output tokens $y_{0:t}$ by the text sequence $x_{1:u}$. For every node, the neural transducer calculates the emission probability~$P(y_{t+1}|u,t)$ and transition probability~$P(\varnothing|u,t)$. The $\langle blank\rangle$ token~$\varnothing$ represents the transition to the next text symbol, indicating the horizontal arrows in Fig.~1. Consequently, the support set of the transducer's output is $ \mathcal{V}\cup \{\varnothing\}$, where $\mathcal{V}$ is the set of the defined semantic tokens. The training objective of the neural transducer is defined as follows: 
\begin{equation} \label{eq1}
\begin{split}
\mathcal{L}_{trans} &= -\log{P(Y|X)} \\
&= -\log{\sum_{a \in \mathcal{F}^{-1}(Y)}{P(a|X)}}.
\end{split}
\end{equation}
In (1), $a$ stands for the possible monotonic alignment and $\mathcal{F}^{-1}$ denotes the inverse function of $\mathcal{F}$. The function $\mathcal{F}$ removes $\varnothing$ from each alignment. Therefore, $\mathcal{L}_{trans}$ is derived as the marginalized likelihood of all the possible monotonic alignment and satisfies $\mathcal{L}_{trans}=\alpha(U,T)P(\varnothing|U,T)$ by the dynamic forward-backward algorithm below:
\begin{equation} \label{eq:loss2}
\begin{split}
  \alpha(u,t) = \alpha(u-1,t)P(\varnothing|u-1,t) \\+ \alpha(u,t-1)P(y_{t}|u,t-1).
\end{split}
\end{equation}
Beyond the basic neural transducer, we use reference speech~$S_{ref}$ for the transducer framework to control the prosody of generated speech. $S_{ref}$ controls the high-level prosody information in semantic tokens, such as speech rate and accent. Therefore, the transducer learns the conditional likelihood of semantic tokens given text and reference speech so that we re-formulate the training objective as follows: 
\begin{equation} \label{eq:loss1}
  \mathcal{L}_{trans'} = -\log{\sum_{a \in \mathcal{F}^{-1}(Y)}{P(a|X, S_{ref})}}.
\end{equation}
Training a neural transducer consumes notoriously large memory because of searching all the possible monotonic paths. To alleviate this problem, we apply a pruning method proposed in \cite{kuang2022pruned}. Excluding the implausible paths, the joiner only computes pruned nodes, which reduces the memory complexity from $O(UT)$ to $O(U)$. Therefore, the final training objective is $\mathcal{L}_{t2t} =  \alpha_{1}\mathcal{L}_{prune} + \alpha_{2}\mathcal{L}_{trans'}$, where $\mathcal{L}_{prune}$ is the loss for searching pruning range and $\{\alpha_{1},\alpha_{2}\}$ are scale factors. Although the pruning method had been explored only for ASR objectives, we found that it also works effectively for our TTS objective. 

\subsubsection{Model architecture}
As described in Fig.~2, the proposed transducer consists of four modules: a text encoder, a reference encoder, a decoder, and a joint net. For the text encoder which takes a phoneme sequence~$x_{1:U}$ and returns text embedding sequence~$h^{enc}_{1:U}$, we leverage conformer blocks~\cite{gulati2020conformer}. The decoder generates token embedding sequence~$h^{dec}_{0:t}$ from the previous tokens~$y_{0:t}$. We use uni-directional LSTM blocks for the decoder. We use LSTM instead of the commonly used transformer block as an AR decoder to reduce the computational cost for fast inference. The reference encoder takes reference speech~$S_{ref}$ and returns a prosodic embedding vector~$h^{ref}$. We exploited the architecture of ECAPA-TDNN~\cite{desplanques2020ecapa} for the reference encoder. Finally, the joint net takes the concatenation of the text embedding and the token embedding~$[h^{enc}_u,h^{dec}_t]$ and returns the categorical distribution of $\mathcal{V}\cup \{\varnothing\}$. The joint net consists of linear layers and conditional layer normalization layers. For conditioning reference speech, the reference embedding~$h^{ref}$ is projected frame-wise for the scale factors of each normalization layer. 

\begin{figure}[]
 \centering
 \includegraphics[width=0.9\columnwidth]{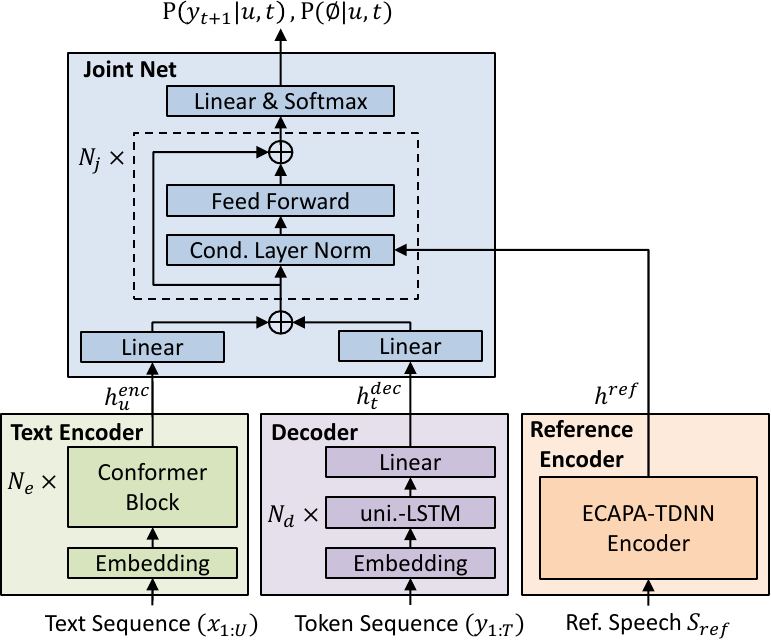}
 \caption{The overall architecture of the proposed neural transducer.}
 \vspace{-0.2cm}
 \label{fig:t-SNE}
 \end{figure}

\subsection{Token-to-Speech~(T2S)}
In the T2S stage, we generate speech from the semantic token. The missing information of the semantic token~(e.g. speaker, recording condition, and acoustic details) is injected by conditioning a reference speech. As semantic tokens are already aligned to speech frames, we don't have to consider the length variability and alignment. Therefore, we design this stage with a NAR architecture rather than AR models for parallel computation. We note that this stage can be trained on audio-only data since semantic tokens are obtained without text labels. It is a significant advantage, as it allows training models on much larger datasets, which can lead to improved performance.

\begin{figure}[]
 \centering
 \includegraphics[width=\columnwidth]{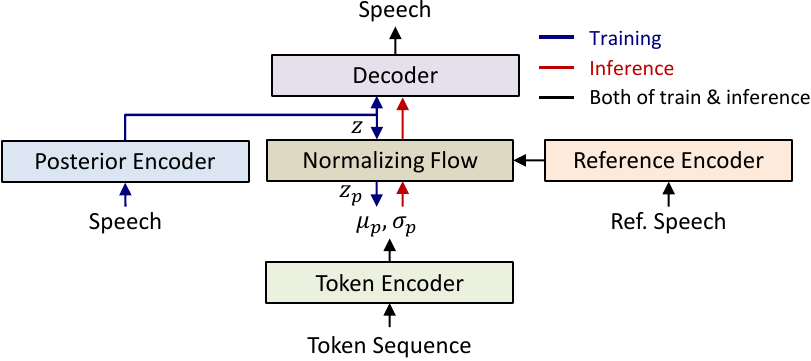}
 \caption{The structure of the VITS-based T2S stage. The colors of the arrows indicate the data flow of the training and inference phases.}
 \vspace{-0.2cm}
 \label{fig:t-SNE}
 \end{figure}

\subsubsection{Model architecture}
We exploit the architecture of VITS~\cite{kim2021conditional} with several modifications. As an end-to-end~(E2E) TTS model, VITS can generate high-quality speech samples with a single training stage. The main modification is that we replace the text input with the semantic token, and accordingly, we replace the text encoder with a token encoder. The token encoder has the same architecture as the posterior encoder of VITS. We do not use the monotonic alignment search~(MAS) algorithm and duration predictor because semantic tokens are already aligned to speech frames. To condition acoustic information, we use a reference encoder with the same structure as that of the T2T stage. The extracted reference embedding conditions affine coupling layers of the normalizing flow as in the original VITS. We clarify that both reference encoders in T2T and T2S stages have the same architecture, but they are trained for different objectives, so that they extract different attributes of speech. We describe the entire architecture of the T2S stage in Fig.~3. We recommend referring to \cite{kim2021conditional, kim2022transfer} for the specific architecture and training objectives.


\section{Experimental Settings}
\label{sec:majhead}

\subsection{Dataset}
We conduct our zero-shot adaptive TTS experiments on the LibriTTS~\cite{zen2019libritts} corpus. For training, we use train subsets of LibriTTS~(train-clean-100, train-clean-360, train-other-500), which contain about 555~hours of labeled speech recorded by 2311~speakers. For evaluation, we use the test-clean subset consisting of about 8~hours of speech with 39~speakers not overlapping with speakers of the training set. The dataset has 24~kHz sample rate, and our experiments are conducted on the same sample rate.

\subsection{Implementation Details}
\textbf{Semantic token:}
We exploit the pre-trained wav2vec2.0-XLSR~\cite{conneau2020unsupervised}\footnote{\url{https://huggingface.co/facebook/wav2vec2-xlsr-53-espeak-cv-ft}} to obtain semantic tokens. We conduct $k$-means clustering on representations of the 15th layer as in \cite{kim2022transfer} and use the cluster index sequence as semantic tokens. We set the number of clusters $k=512$ for our experiments.\\
\textbf{T2T implementation:}
We use the icefall toolkit\footnote{\url{https://github.com/k2-fsa/icefall}} to implement the neural transducer. We convert text sequences to International Phonetic Alphabet~(IPA) sequences using phonemizer library\footnote{\url{https://github.com/bootphon/phonemizer}}. For the text encoder, we build 6~conformer blocks\footnote{We used the modified version of conformer not officially documented. We recommend referring the code for the detailed architecture:\\ \url{https://github.com/k2-fsa/icefall/blob/master/egs/librispeech/ASR/pruned_transducer_stateless6/conformer.py}} with 384~hidden dimension, 1536~feed forward dimension, and kernel size~3. The decoder consists of 2~layers of uni-directional LSTM with 512~hidden dimension. The joint net has 3~feed forward blocks with 512~hidden dimension. As a reference speech for training, we use the target speech randomly sliced to be 3~seconds. We found that slicing the reference speech reduces the mismatch between training and inference and improve the pronunciation. For the pruned training method, we verify that the pruning bounds constant~$S=50$ operates well for our experiments. The model is trained for 30 epochs with a dynamic batch size containing up to 240~seconds of speech per iteration. We use Adam optimizer with $\beta_1=0.9$, $\beta_2 =0.98$, and learning rate 0.05. The training takes about 54~hours on 2~Quadro RTX8000 GPUs.\\
\textbf{T2S implementation:}
We follow the configuration of \cite{kim2021conditional} unless otherwise explained. To match the sample rate to be 24~kHz, we adjust the intermediate frame length to be 10ms, 
and upsampling rates of the decoder to be $(10, 6, 2, 2)$. For efficient training, we randomly slice each sample to be 2~seconds at every iteration. We train the model for 400k iterations with batch size 64, which takes about 5~days on 2 Quadro RTX8000 GPUs.


\subsection{Baselines}
We employ two baseline models for comparison: VITS and VALL-E~\cite{wang2023neural}. To build the baseline VITS on a zero-shot adaptive scenario, we use an ECAPA-TDNN as a reference encoder. Thus, the baseline VITS has the same architecture as our T2S model, except for the text encoder, duration predictor, and the monotonic alignment search~(MAS) algorithm. The baseline VITS is used to represent the effectiveness of using semantic tokens generated by our neural transducer compared to directly using text input. We adopted the transfer learning method proposed in \cite{kim2022transfer} for training stability, exploiting our T2S model as a pre-trained model. 

For the baseline VALL-E, we use an unofficial implementation\footnote{\url{https://github.com/lifeiteng/vall-e}}. The baseline VALL-E is trained on the same dataset, LibriTTS, for about 4~days on 8~Tesla A100 GPUs. We clarify that the baseline VALL-E shows degraded performance compared to the paper because the unofficial implementation is trained on a much smaller training dataset and resources.

\section{Results}

\subsection{Overall Performance}
We evaluate the generated sample quality on subjective and objective measures. For subjective evaluation, we conducted a mean opinion score~(MOS) and a similarity mean opinion score~(SMOS) test. In the MOS test, 16 testers rated the quality of randomly selected samples with 5~scaled scores from 1 to 5 regarding naturalness and intelligibility. In the SMOS test, the same testers listened to the pairs of the reference and generated speech, and rated the similarity of speaker and prosody in 5~scale. For objective measures, we calculate character error rate~(CER) and speaker embedding cosine similarity~(SECS). CER represents the intelligibility of samples using an ASR model. We transcribed samples using the whisper-large model~\cite{radford2022robust}. To calculate the speaker similarity, we use a pre-trained speaker verification model from speechbrain~\cite{ravanelli2021speechbrain}. The SECS is ranged from  -1 to 1, where the higher score means the higher similarity. The results of all of the measures are presented in Table~1. 

\begin{table}[th]
\setlength{\tabcolsep}{5pt}
\setlength{\arrayrulewidth}{0.2mm}
\caption{Results of zero-shot adaptive TTS. MOS and SMOS are represented with 95\% confidence intervals.}
\vspace{-0.2cm}
\label{tab:Model size}
\centering
\begin{tabular}{l c c c c}
\toprule
\textbf{Method} & \textbf{MOS}& \textbf{SMOS} & \textbf{CER(\%)}& \textbf{SECS} \\ 
\midrule
Ground Truth & 4.62\footnotesize{$\pm$0.06} & 4.41\footnotesize{$\pm$0.08} & 1.02 & 0.673   \\
\midrule
VITS      &3.29\footnotesize{$\pm$0.08} & 3.89\footnotesize{$\pm$0.08} & 11.19 & 0.492                             \\
VALL-E     &3.47\footnotesize{$\pm$0.09} & 3.59\footnotesize{$\pm$0.09}  & 21.67 & 0.350   \\
Proposed     &\textbf{4.11\footnotesize{$\pm$0.07}} &\textbf{4.42\footnotesize{$\pm$0.06}}   & \textbf{3.05} & \textbf{0.504}      \\
\bottomrule
\end{tabular}
\end{table}

As shown in Table~1, the proposed model shows higher performance in all the metrics. Compared to VITS, our performance improvements mainly stems from using intermediate semantic tokens. The results also verify that the neural transducer is a good candidate for semantic token generation. Although VALL-E generally generates high-fidelity samples, we notice that VALL-E sometimes generates misaligned samples with skipping or repetition, which critically degraded the overall performance. We assume that it is because the VALL-E doesn't have structural constraints on monotonic alignment and fully relies on self-attention. On the contrary, the proposed model shows more robust and stable performance thanks to the monotonic alignment constraints of the neural transducer. One notable point is that the proposed model shows comparable SMOS to the ground truth. It is because even if the proposed model lacks speaker similarity compared to the ground truth, the proposed model also reflects the prosody of reference speech, whereas the ground truth only shares speakers with the reference speech.

\subsection{Inference Speed}
We evaluate the inference speed on a Quadro RTX8000 GPU. We fix the reference speech, which corresponds to the prompt for VALL-E, to a 5-second sample. The result is shown in Table~2 and Fig.~4. Overall, VITS shows the fastest inference speed with a large margin because of its fully NAR architecture. On the contrary, VALL-E shows the slowest inference speed, which is due to its transformer-based AR structure and relatively large model size. The proposed model has a slower inference speed than VITS but is much faster than real-time. This is because the proposed model has a hybrid architecture that combines an AR T2T stage and a NAR T2S stage. Although the transducer has an AR decoding and linearly increasing tendency, the small LSTM-based decoder makes our model work well enough in real-time.
 
\begin{table}[th]
\setlength{\tabcolsep}{8pt}
\setlength{\arrayrulewidth}{0.2mm}
\caption{Inference speed on a Quadro RTX8000 GPU.}
\vspace{-0.2cm}
\label{tab:Model size}
\centering
\begin{tabular}{l c c c c}
\toprule
\textbf{Method} & \textbf{VITS}& \textbf{VALL-E}& \textbf{Proposed}\\ 
\midrule
$\times$Real-time        & 96.40 & 0.314 & 10.28               \\
\bottomrule
\end{tabular}
\end{table}

\begin{figure}[th]
 \centering
 \includegraphics[width=0.9\columnwidth]{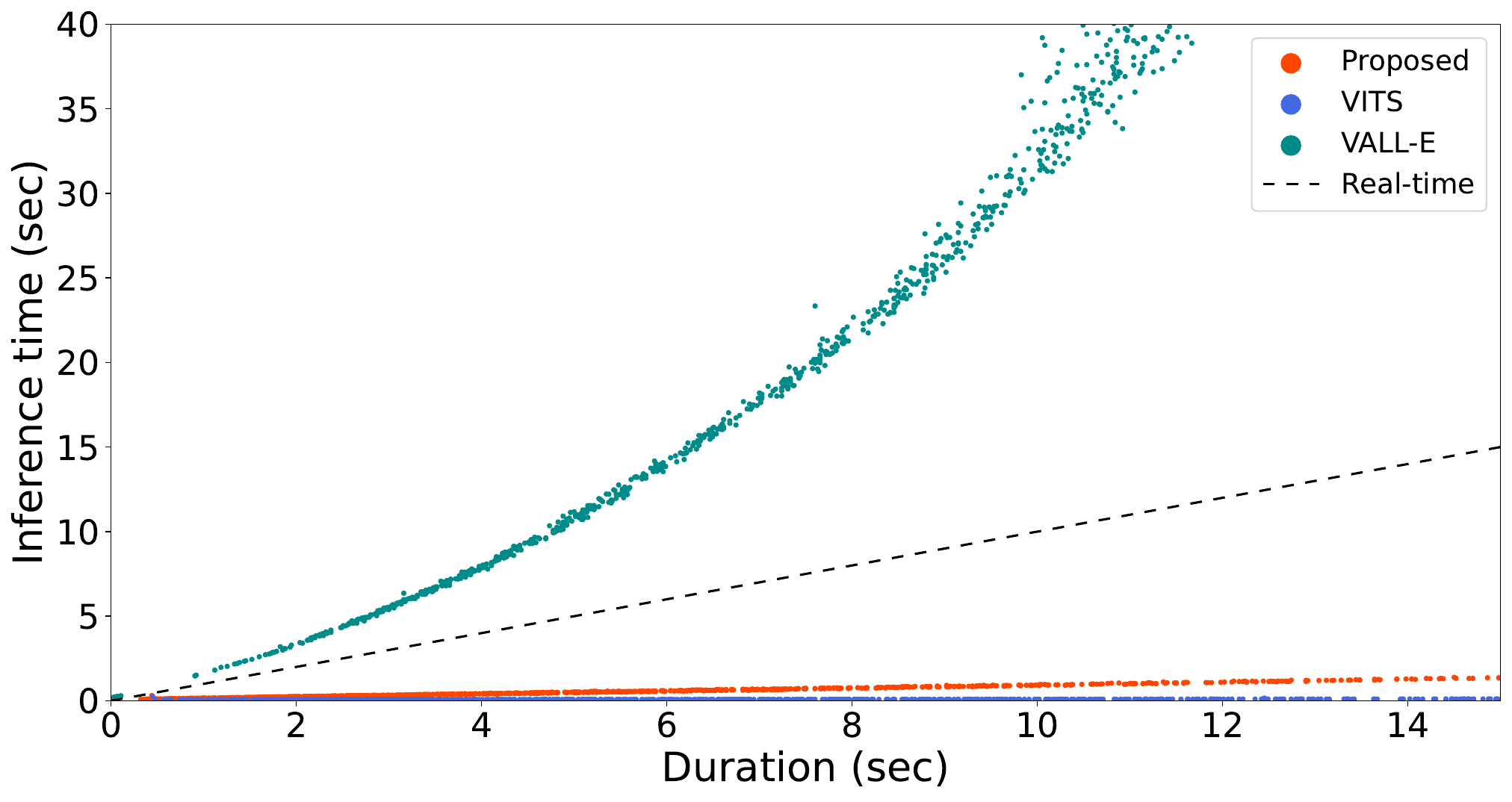}
 \caption{Visualization of inference speed.}
 \vspace{-0.2cm}
 \label{fig:t-SNE}
 \end{figure}

\subsection{Controllability of Neural Transducer}
We verify the prosody controllability of our neural transducer by comparing the mel-spectrograms of the reference speech and the generated speech. We fix the text and reference speech for the T2S stage, and allowed only the neural transducer to control the prosody of the generated speech. As shown in Fig.~5, the generated samples show various prosody depending on the given reference speech. In particular, duration-related attributes such as silence between segments and speech rate are well reflected in the generated speech. We calculate the speech rate as $\frac{the\ number\ of\ phonemes}{duration\ of\ speech~(sec)}$, and show the positive correlation of speech rate between the reference and generated speech in Fig.~5. It demonstrates that our neural transducer can control the speaking speed using the reference speech. We upload audio samples on our demo page to demonstrate more detailed results.


\begin{figure}[th]
 \centering
 \includegraphics[width=0.9\columnwidth]{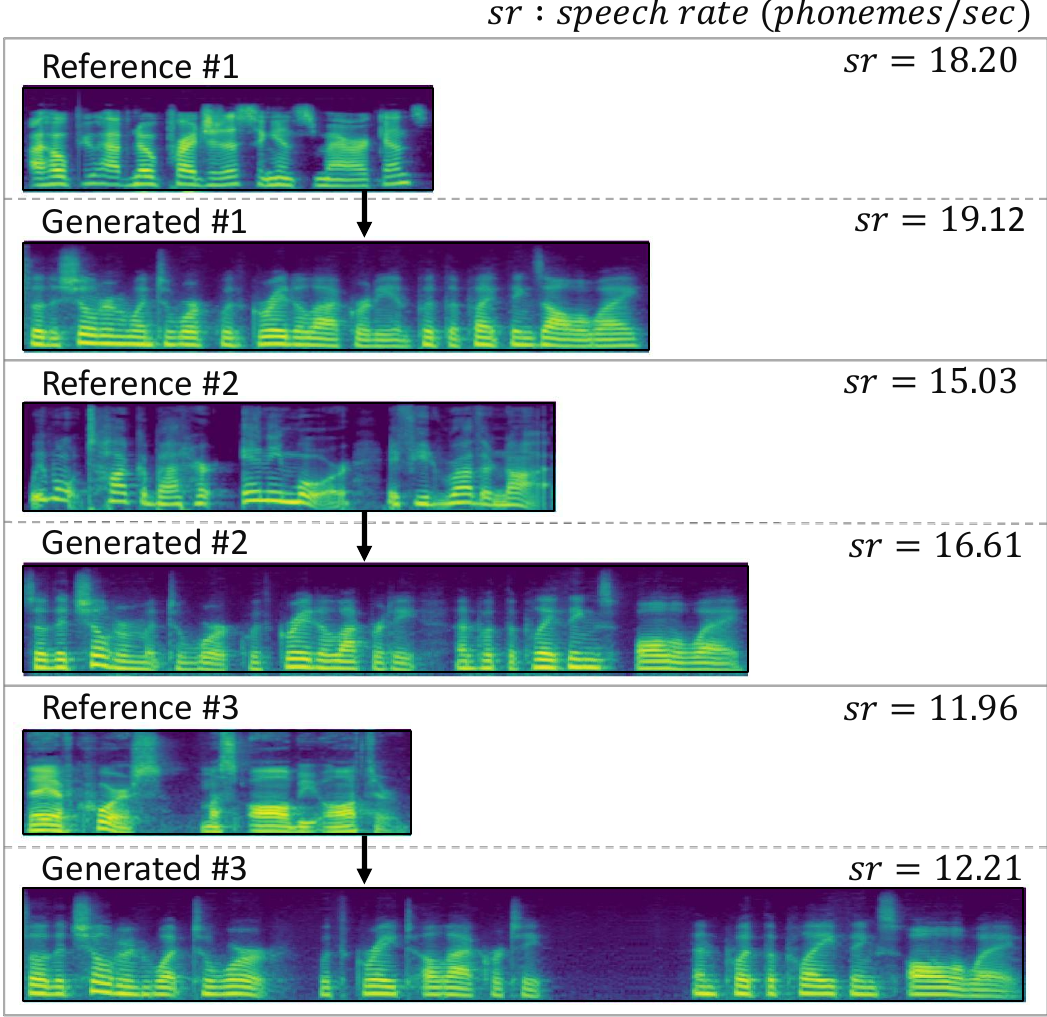}
 \caption{Mel-spectrogram visualization for pairs of reference speech~(T2T) and generated speech.}
 \vspace{-0.2cm}
 \label{fig:t-SNE}
 \end{figure}

\section{Conclusion}
\label{sec:ref}
We proposed a novel TTS framework that uses a neural transducer to predict semantic tokens. The semantic tokens are highly disentangled from complex acoustic variability, thanks to wav2vec2.0 embeddings. This allows the neural transducer to easily learn the alignment and semantic information, taking advantage of the monotonic alignment constraints. A VITS-based speech generator then synthesizes speech from semantic tokens, focusing on the acoustic details. Our experimental results show that the proposed two-stage framework outperforms the baselines regarding speech quality and zero-shot adaptation. In addition, it also has benefits in inference speed and prosody controllability. In future work, we will further investigate the neural transducer for TTS systems regarding architecture improvement and controllability.

\bibliographystyle{IEEEbib}
\bibliography{strings,refs}

\end{document}